\documentstyle[twoside,fleqn,espcrc2]{article}

\newcommand{\AmS}{{\protect\the\textfont2
  A\kern-.1667em\lower.5ex\hbox{M}\kern-.125emS}}

\title{\sf Lorentz symmetry violation and high-energy cosmic rays}

\author{\sf L. Gonzalez-Mestres  
\address{\sf Laboratoire de Physique Corpusculaire, 
        Coll\`ege de France, 75231 Paris 
Cedex 05 , France}$^,$\address{\sf Laboratoire  
d'Annecy-le-Vieux de Physique des Particules, 
B.P. 110 , 74941 Annecy-le-Vieux Cedex}}
       
\begin{document}

\begin{abstract}
We discuss possible violations of Poincar\'e's relativity principle 
at energy scales close to Planck scale and
point out the potentialities of high-energy cosmic-ray physics 
to uncover these new phenomena.
\end{abstract}

\maketitle
{\sf
\section{THE RELATIVITY PRINCIPLE: A BASIC PHYSICS ISSUE BEHIND 
A DEBATE ON PRIORITY}

H. Poincar\'e was the first author to consistently 
formulate the relativity principle stating in 
1895 \cite{Poincare95}:

{\it  
"Absolute motion of matter, or, to be more precise, the
relative motion of weighable matter and ether, cannot be disclosed. All that
can be done is to reveal the motion of weighable matter with respect to
weighable matter"}

The deep meaning of this law of Nature was further
emphasized when he wrote \cite{Poincare01}:

{\it "This principle will be confirmed with increasing precision, 
as measurements 
become more and more accurate"}

The role of H. Poincar\'e in building relativity, 
and the relevance of his thought, have 
often been emphasized \cite{Logunov95,Fey}. 
In his
June 1905 paper \cite{Poincare05}, published before Einsteins's article
\cite{Einstein05} arrived (on June 30) to the editor,
he explicitly wrote the relativistic transformation law for the 
charge density and velocity of motion and applied to gravity
the "Lorentz group", assumed to hold for "forces of whatever origin". 
However, his priority is sometimes 
denied \cite{Miller96,Paty96} on the grounds that {\it "Einstein essentially
announced the failure of all ether-drift experiments past and future as a 
foregone conclusion, contrary to Poincar\'e's empirical bias"} \cite{Miller96},
that Poincar\'e did never {\it "disavow the ether"} \cite{Miller96} or that
{\it "Poincar\'e never challenges... the absolute time of newtonian 
mechanics... the ether is not only the absolute space of mechanics... but a 
dynamical entity"} \cite{Paty96}. It is implicitly assumed that A. Einstein
was right in 1905 when {\it "reducing ether to the absolute space of 
mechanics"} \cite{Paty96} and that H. Poincar\'e was wrong because {\it "the
ether fits quite nicely into Poincar\'e's view of physical reality: the ether 
is real..."} \cite{Miller96}. But modern particle physics has 
brought back the concept of a non-empty vacuum where free particles propagate:
without such an "ether" where fields can condense, the standard model 
of electroweak interactions could not be written and quark confinement could
not be understood. Modern cosmology is not incompatible with
an "absolute local frame" close to that suggested by the study of cosmic
microwave background radiation. Then, the relativity principle would become
a symmetry, a concept whose paternity was attributed to
H. Poincar\'e by R.P. Feynman \cite{Feynman}:

{\it "Precisely Poincar\'e proposed investigating what could be done with the
equations without altering their form. It was precisely his idea to pay
attention to the symmetry properties of the laws of physics"}

As symmetries in particle physics are in general violated, 
Lorentz symmetry may be broken and an absolute local rest frame may 
be detectable through experiments performed beyond the relevant scale.
Poincar\'e's special relativity (a symmetry applying to
physical processes) could live with this situation, but not Einstein's 
approach such as it was formulated in 1905 (an absolute geometry of space-time
that matter cannot escape). But, is Lorentz symmetry broken? 
We discuss here two issues: a) the scale 
where we may expect Lorentz symmetry to be violated; b) the physical phenomena
and experiments potentially able to uncover Lorentz symmetry violation (LSV).
Previous papers on the subject are references \cite{Gon96} to \cite{Gon9710}
and references therein. 

\section{SPECIAL RELATIVITY AS A LOW-ENERGY LIMIT}

Low-energy tests of special relativity have cosfirmed its validity to an
extremely good accuracy, but the situation at very high energy remains more
uncertain (see \cite{Gon96} to \cite{Gon9710}). 
If Lorentz symmetry violation (LSV) follows
a $E^2$ law 
($E$ = energy), similar to the effective gravitational coupling, it can
be $\approx ~1$ at $E~\approx ~10^{21}~eV$ and $\approx ~10^{-26}$ at $E~
\approx ~100~MeV$ (corresponding to the highest momentum scale involved in
nuclear magnetic resonance experiments), in which case it will escape all
existing low-energy bounds. If LSV is $\approx ~1$ at Planck scale ($E~
\approx ~10^{28}~eV$), and following a similar law, it will be $\approx 
~10^{-40}$ at $E~\approx ~100~MeV$ . Our suggestion is not in contradiction
with Einstein's thought such as it became after he had developed general
relativity. In 1921 , A. Einstein wrote in "Geometry and Experiment":
{\it "The interpretation of geometry advocated here cannot be directly applied
to submolecular spaces... it might turn out that such an extrapolation is
just as incorrect as an extension of the concept of temperature to particles
of a solid of molecular dimensions"}.
It is remarkable that special relativity holds at the
attained accelerator energies, but there is no fundamental reason for this to
be the case above Planck scale. 

A typical example of models violating Lorentz symmetry at very short distance
is provided by models where an absolute local rest frame exists and
non-locality in space is introduced through a
fundamental length scale $a$ \cite{Gon9703}. Such models lead to a deformed
relativistic kinematics of the form \cite {Gon9703,Gon9710}:
\equation
E~=~~(2\pi )^{-1}~h~c~a^{-1}~e~(k~a)
\endequation
\noindent
where $h$ is the Planck constant, $c$ the speed of light, $k$ the wave vector
and
$[e~(k~a)]^2$ is a convex
function of $(k~a)^2$ obtained from vacuum dynamics.
Expanding equation (1) for $k~a~\ll ~1$ , we can write:
\begin{eqnarray}
e~(k~a) & \simeq & [(k~a)^2~-~\alpha ~(k~a)^4 \nonumber 
\\
& &~~~~+~(2\pi ~a)^2~h^{-2}~m^2~c^2]^{1/2}
\end{eqnarray}
\noindent
$m$ being the mass, $\alpha $ a model-dependent constant 
$\approx 0.1~-~0.01$ for
full-strength violation of Lorentz symmetry at  
momentum scale $p~\approx ~ a^{-1}~h~$,
and:
\begin{eqnarray}
E & \simeq & p~c~+~m^2~c^3~(2~p)^{-1} \nonumber
\\
& &~~~~~~~~~~~~~~~-~p~c~\alpha ~(k~a)^2/2~~~~~
\end{eqnarray}
The "deformation" $\Delta ~E~=~-~p~c~\alpha ~(k~a)^2/2$ in the right-hand
side of (3) implies a Lorentz symmetry violation in the ratio $E~p^{-1}$
varying like $\Gamma ~(k)~\simeq ~\Gamma _0~k^2$ where $\Gamma _0~
~=~-~\alpha ~a^2/2$ . If $c$ and $\alpha $ are universal parameters for all
particles, LSV does not lead to the spontaneous decays predicted in
\cite{Glashow}: the existence of very high-energy cosmic rays 
cannot be regarded as an evidence against LSV. With the deformed relativistic
kinematics (DRK) defined by (1)-(3), Lorentz symmetry remains valid 
in the limit $k~\rightarrow ~0$, contrary to the standard $TH\epsilon \mu $
model \cite{Will}. The above non-locality may actually be an approximation to
an underlying dynamics involving superluminal particles 
\cite{Gon96,Gon9705,Gon9709,Gon9710}, just as electromagnetism looks nonlocal 
in the potential approximation to lattice dynamics in solid-state physics:
it would then correspond to the limit $c~c_i^{-1}~\rightarrow~0$ 
where $c_i$ is the superluminal critical speed.

Are $c$ and $\alpha $ universal? This may be 
the case for all "elementary" particles, i.e. 
quarks, leptons, gauge bosons...,
but the situation is less obvious for hadrons, nuclei and heavier objects.
From a naive soliton model \cite{Gon9703}, we inferred that: a) $c$ is 
expected to be universal up to very small corrections ($\sim 10^{-40}$) 
escaping all existing bounds; b)
an  
approximate rule can be to take $\alpha $ universal for leptons, gauge bosons
and light hadrons (pions, nucleons...) and assume a $\alpha \propto m^{-2}$
law for nuclei and heavier objects, the nucleon mass setting the scale. 

\section{THE RELEVANCE OF COSMIC-RAY EXPERIMENTS}

If Lorentz symmetry is broken at Planck scale or at some other 
fundamental length scale, the effects of LSV may be observable well below this
energy: they can produce detectable phenomena at the highest
observed cosmic ray energies. This is due to DRK 
\cite{Gon9703,Gon9706,Gon9707}: at energies above 
$E_{trans}~
\approx ~\pi ^{-1/2}~ h^{1/2}~(2~\alpha )^{-1/4}~a^{-1/2}~m^{1/2}~c^{3/2}$, 
the very small deformation $\Delta ~E$ 
dominates over
the mass term $m^2~c^3~(2~p)^{-1}$ in (3) and modifies all
kinematical balances. Because of the negative value of $\Delta ~E$ , it costs 
more and more energy, as energy increases above $E_{trans}$, 
to split the incoming logitudinal momentum.
The parton model (in any version), as well as standard
formulae for Lorentz contraction and time dilation, are also expected to fail
above this energy \cite{Gon9703,Gon9710} which corresponds to $E
~\approx~10^{20}~eV$
for $m$ = proton mass and
$\alpha ~a^2~\approx ~10^{-72}~cm^2$ (f.i. $\alpha 
~\approx ~10^{-6}$ and $a$ = Planck
length), and to $E~\approx ~10^{18}~eV$ for
$m$ = pion mass and $\alpha ~a^2~\approx ~10^{-67}~cm^2$
(f.i. $\alpha ~\approx ~0.1$ and $a$ = Planck length). 
Assuming that the earth moves slowly with
respect to the absolute rest frame
(the "vacuum rest frame"), these 
effects lead to observable phenomena 
in future experiments devoted to the highest-energy cosmic-rays:

a) For $\alpha ~a^2~>~10^{-72}~cm^2$ , and
assuming a universal value of $\alpha $ ,
there is no Greisen-Zatsepin-Kuzmin
cutoff for the particles under 
consideration and cosmic rays (e.g. protons) 
from anywhere in the presently observable Universe
can reach the earth.

b) With the same hypothesis, 
unstable particles with at
least two stable particles in the final states   
of all their decay channels become stable at very 
high energy. Above $E_{trans}$, the lifetimes of all
unstable particles (e.g. the $\pi ^0$ in 
cascades) become much longer than predicted
by relativistic kinematics.

c) In astrophysical processes at very
high energy, 
similar mechanisms can inhibit radiation under 
external forces, GZK-like cutoffs, decays, 
photodisintegration of nuclei, momentum loss through
collisions, production of lower-energy secondaries...
potentially contributing to solve all basic problems
raised by the highest-energy cosmic rays.

d) With the same hypothesis, the allowed final-state 
phase space of two-body collisions is modified and
can lead to a sharp fall of cross-sections 
for incoming cosmic ray energies above
$E_{lim} ~\approx ~(2~\pi )^{-2/3}~(E_T~a^{-2}~ \alpha ^{-1}~h^2~c^2)^{1/3}$,
where $E_T$ is the energy of the target. As a consequence, and with the
previous figures for Lorentz symmetry violation, above some
energy $E_{lim}$ between 10$^{22}$ and $10^{24}$ $eV$ a cosmic  
ray will not deposit most of its energy in the atmosphere
and can possibly fake an exotic event with much less energy. 

e) Effects a) to d) are obtained using only DRK. If dynamical
anomalies are added (failure, at very small distance scales, 
of the parton model and of the
standard Lorentz formulae for length and time...), we can expect 
much stronger effects in the cascade development profiles
of cosmic-ray events.

f) Cosmic superluminal particles would produce atypical events 
with very small total momentum, isotropic or involving several
jets \cite{Gon96,Gon9705,Gon9709,Gon9710}.

}
\end{document}